\pdfoutput=1
\documentclass{article}

\usepackage{arxiv}
\usepackage[utf8]{inputenc}
\usepackage[T1]{fontenc}
\usepackage{microtype}
\usepackage{amsmath,amssymb,amsfonts}
\usepackage{graphicx}
\usepackage{booktabs}
\usepackage{float}
\usepackage[font=small,labelfont=bf]{caption}
\usepackage{placeins}
\usepackage[numbers,sort&compress]{natbib}
\usepackage{url}
\usepackage[colorlinks=true,allcolors=blue]{hyperref}
\hypersetup{pdftitle={Radical-Fragment Many-Body Expansion for Linear Alkane Quantum Chemistry}, pdfauthor={Daniel Sierra-Sosa, Jorge Saavedra, Santiago Solares, Gregorio Toscano-Pulido}, pdfkeywords={quantum chemistry, many-body expansion, fragment molecular orbital, variational quantum eigensolver, ADAPT-VQE, sample-based quantum diagonalization, linear alkanes, radical}}
\usepackage[capitalise,nameinlink]{cleveref}

\newcommand{\ket}[1]{\left|#1\right\rangle}
\newcommand{\bra}[1]{\left\langle#1\right|}

\newcommand{\expect}[1]{\left\langle #1 \right\rangle}
\title{Radical-Fragment Many-Body Expansion for Linear Alkane Quantum Chemistry}

\author{%
Daniel Sierra-Sosa\\
Department of Computer Science\\
The Catholic University of America\\
\texttt{sierrasosa@cua.edu}
\And
Jorge Saavedra\\
Department of Computer Science\\
The Catholic University of America\\
\texttt{saavedrajor@cua.edu}
\AND
Santiago Solares\\
Department of Mechanical Engineering\\
The Catholic University of America\\
\texttt{solares@cua.edu}
\And
Gregorio Toscano-Pulido\\
Department of Computer Science\\
The Catholic University of America\\
\texttt{gtoscano@cua.edu}
}
\date{}
\renewcommand{\headeright}{Preprint}
\renewcommand{\undertitle}{Preprint}
\renewcommand{\shorttitle}{Radical-Fragment MBE2 for Linear Alkanes}

\begin{document}
\maketitle

\begin{abstract}
We introduce a radical-fragment many-body expansion at the two-body level (MBE2) for quantum chemistry of linear alkanes. Instead of heterolytic bond cleavage with hydrogen capping atoms and electrostatic embedding like in Fragment Molecular Orbital (FMO), we perform homolytic C--C bond cleavage to produce open-shell radical fragments ($\mathrm{CH}_3^{\bullet}$, ${}^{\bullet}\mathrm{CH}_2^{\bullet}$) treated with restricted open-shell Hartree--Fock (ROHF) in isolation. The two-body MBE2 assembly formula reconstructs total alkane energies from only four unique fragment calculations regardless of chain length, reducing the maximum qubit requirement. We benchmark this framework against five energy solvers (RHF, CCSD, VQE, ADAPT-VQE, and SQD) across 11 linear alkanes from butane ($\mathrm{C}_4\mathrm{H}_{10}$) to hexacosane ($\mathrm{C}_{26}\mathrm{H}_{54}$). The MBE2 decomposition achieves a 12.3$\times$ qubit reduction for $\mathrm{C}_{26}\mathrm{H}_{54}$ (from 368 to 30 qubits) and a 12.8$\times$ reduction in unique calculations via symmetry exploitation. MBE2-VQE and MBE2-SQD (executed on IBM quantum hardware) closely track their respective classical MBE2 references, demonstrating that fragmentation-based quantum chemistry is viable for scaling quantum solvers to large molecular systems.
\end{abstract}

\keywords{quantum chemistry \and many-body expansion \and fragment molecular orbital \and variational quantum eigensolver \and ADAPT-VQE \and sample-based quantum diagonalization \and linear alkanes \and radical}

\section{Introduction}\label{sec:introduction}

Linear alkanes $\mathrm{C}_n\mathrm{H}_{2n+2}$ are among the most fundamental
organic molecules in chemistry. Their simple, repetitive backbone of
sp$^3$-hybridized carbon atoms connected by C--C $\sigma$-bonds makes them ideal
testbeds for electronic structure methods. Despite their structural
simplicity, alkanes pose severe computational challenges as chain length
grows. The number of Slater determinants spanning the full configuration
interaction (FCI) space grows combinatorially with the number of
electrons and orbitals, making exact solutions intractable beyond small
systems. Under the Jordan--Wigner transformation with the STO-3G minimal
basis, each alkane requires $q=14n+4$ qubits, yielding 368 qubits for
hexacosane, far beyond current quantum hardware capabilities \cite{kitaura1999fmo}.

The repeating, symmetric structure of linear alkanes, where every
internal $\mathrm{CH}_2$ group is nearly chemically equivalent and every
terminal $\mathrm{CH}_3$ group is nearly equivalent, creates a natural
opportunity for fragmentation strategies that decompose the problem into
small, constant-size subproblems. The standard Fragment Molecular
Orbital method \cite{kitaura1999fmo,fedorov2007extending,fedorov2017fragment} achieves this through either homolytic or
heterolytic bond cleavage, as needed, with hydrogen capping atoms,
producing closed-shell fragments computed in an electrostatic embedding
potential with self-consistent charge (SCC) iterations.

We propose a different strategy based on a radical-fragment many-body
expansion (MBE2) via homolytic C--C bond cleavage. We split each bonding
electron pair symmetrically, producing open-shell radical species:
methyl radical $\mathrm{CH}_3^{\bullet}$ (doublet) and methylene diradical
${}^{\bullet}\mathrm{CH}_2^{\bullet}$ (open-shell), which are
left uncapped and are treated with restricted open-shell Hartree-Fock
(ROHF) without electrostatic embedding. This homolytic fragmentation is
chemically natural for non-polar C--C bonds where bond dissociation is
approximately symmetric.

This paper is organized as follows. Section~\ref{sec:methodology} presents the theoretical
foundations of the radical-fragment MBE2 decomposition, the homolytic
cleavage scheme, and three quantum solver pipelines: Variational Quantum
Eigensolver (VQE) \cite{peruzzo2014vqe}, Adaptive Derivative-Assembled Pseudo-Trotter
ansatz Variational Quantum Eigensolver (ADAPT-VQE) \cite{grimsley2019adapt}, and
Sample-based quantum diagonalization (SQD) on IBM quantum hardware
\cite{robledoMoreno2025sqd}, alongside the classical Restricted Hartree-Fock (RHF) and
Cluster Singles and Doubles (CCSD) references. Section~\ref{sec:computational-details} describes the
computational setup, including geometry construction, basis set, and
solver-specific parameters. Section~\ref{sec:results} benchmarks all energy evaluation
strategies across 11 linear alkanes from butane ($\mathrm{C}_4\mathrm{H}_{10}$) to hexacosane
($\mathrm{C}_{26}\mathrm{H}_{54}$), demonstrating a 12.3$\times$ qubit reduction and close agreement
between the quantum solvers and their classical MBE2 counterparts.
Section~\ref{sec:discussion} discusses implications, and Section~\ref{sec:conclusions} concludes with an
outlook on future extensions.

\section{Methodology}\label{sec:methodology}

The standard Fragment Molecular Orbital (FMO) method, developed by
Kitaura et al. (1999) \cite{kitaura1999fmo}, decomposes large molecules into fragments
that are computed in the electrostatic embedding potential of the entire
system, with self-consistent charge (SCC) iterations between fragments.
When covalent bonds are cut, FMO typically uses bond-detached atoms
(BDA) or hydrogen-cap atoms to saturate dangling bonds, producing
closed-shell fragments. The electrostatic embedding and SCC iterations
are essential to the FMO formalism and distinguish it from simpler
fragmentation approaches.

Our Radical-Fragment Many-Body Expansion (MBE2) approach performs
homolytic C--C bond cleavage, splitting each bonding electron pair
symmetrically so that every fragment retains one electron per cut bond.
This produces electrically neutral, open-shell radical species which are
treated with Restricted Open-Shell Hartree--Fock (ROHF) as isolated
systems, without electrostatic embedding, self-consistent charge
iterations, or capping atoms. The bonded-pair dimers also retain their
open-shell character: the $\mathrm{CH}_3-\mathrm{CH}_2$ dimer is a 17-electron
doublet ($S=1/2$) and the $\mathrm{CH}_2-\mathrm{CH}_2$ dimer is a 16-electron
triplet ($S=1$), so the unpaired electrons are never artificially
re-paired upon fragment recombination. For the non-polar C--C bonds of
linear alkanes, where both bonded carbons have identical
electronegativity ($\Delta\chi=0$), homolytic cleavage is the
physically natural description of bond dissociation, and the resulting
radical fragments preserve the local electronic structure of the intact
molecule more faithfully than heterolytic alternatives that introduce
artificial charges or saturating atoms.

\subsection{Many-Body Expansion Assembly}\label{subsec:mbe2}

The many-body expansion (MBE) is a general framework for expressing the
total energy of a composite system as a hierarchy of fragment
contributions \cite{fedorov2007extending,fedorov2017fragment}. At the two-body level (MBE2), the total energy
is reconstructed from monomer energies and pairwise bonded-dimer
corrections:

\begin{equation}\label{eq:mbe2}
E_{\mathrm{MBE2}} = \sum_I E_I + \sum_{I<J}^{\mathrm{bonded}} \Delta E_{IJ}.
\end{equation}

where \(E_{I}\) denotes the energy of radical monomer fragment \(I\) and
\(E_{IJ}\) is the energy of bonded dimer pair \(IJ\). The pair
correction $\Delta E_{IJ}=E_{IJ}-E_I-E_J$ captures the bonding
interaction energy that is absent in the isolated radical monomers. The
bonded-pair-only restriction means we include corrections only for
adjacent fragments \((I,\ I + 1)\) along the chain, neglecting
through-space interactions between non-bonded fragments.

For linear alkanes with translational symmetry, every internal
$\mathrm{CH}_2$ group is chemically nearly equivalent and every terminal
$\mathrm{CH}_3$ group is also nearly equivalent, so (1) can be reduced to
only four unique fragment calculations regardless of chain length---two
monomer types ($\mathrm{CH}_3^{\bullet}$,
${}^{\bullet}\mathrm{CH}_2^{\bullet}$) and two bonded-dimer types
($\mathrm{CH}_3-\mathrm{CH}_2$, $\mathrm{CH}_2-\mathrm{CH}_2$). Expanding (1) for a chain
with \emph{n} carbon atoms gives:

\begin{equation}\label{eq:mbe2-linear}
E_{\mathrm{MBE2}}(n) = 2E_{\mathrm{CH}_3^{\bullet}} + (n - 2)E_{{}^{\bullet}\mathrm{CH}_2^{\bullet}} + 2\Delta E_{\mathrm{CH}_3-\mathrm{CH}_2} + (n - 3)\Delta E_{\mathrm{CH}_2-\mathrm{CH}_2}.
\end{equation}

This constant-cost property is the central practical advantage:
regardless of whether the alkane has 4 or 26 carbons, only 4 unique
quantum calculations are required, and the maximum size of the
calculation (30 qubits for the $\mathrm{CH}_3-\mathrm{CH}_2$ dimer under
Jordan--Wigner mapping with STO-3G) remains fixed. The key distinction
from standard FMO is that our monomers are open-shell radical species
computed without electrostatic embedding, while standard FMO monomers
are closed-shell fragments embedded in the system's electrostatic
potential. The subsections below detail this distinction and describe
each energy solver used to evaluate the fragment energies in (1).

\subsection{Classical Methods: RHF and CCSD}\label{subsec:classical}

In this work, we evaluate the proposed method using Restricted
Hartree-Fock and Coupled Cluster Singles and Doubles as a baseline. RHF
provides the zeroth-order mean-field description of each fragment
through a single-determinant approximation, capturing the dominant
contribution to the total energy at low computational cost. However,
because it neglects electron correlation beyond the average
electron-electron interaction, RHF alone is insufficient for chemically
accurate predictions. CCSD addresses this limitation by introducing
single and double excitations on top of the Hartree-Fock reference,
yielding a more accurate description of dynamical correlation.

\subsubsection{Restricted Hartree-Fock (RHF)}\label{subsubsec:rhf}

The Hartree-Fock method approximates the many-electron wavefunction by a
single Slater determinant \(\ket{\Phi_0}\), built from
a set of molecular orbitals \(\{\phi_i\}\). Within the
restricted formalism, the total electronic energy is obtained
variationally as:

\begin{equation}\label{eq:rhf-energy}
E_{\mathrm{RHF}} = \sum_i^{\mathrm{occ}} h_{ii} + \frac{1}{2}\sum_{i,j}^{\mathrm{occ}}\left(J_{ij}-K_{ij}\right) + V_{NN}.
\end{equation}

where
\(h_{ii} = \ \langle\phi_{i}| - \frac{1}{2}\nabla^{2} - \sum_{A}^{}\frac{Z_{A}}{R_{iA}}|\phi_{i}\rangle\)
are the one-electron integrals,
\(J_{ij} = \langle\phi_{i}\phi_{j}\ |\ \phi_{i}\phi_{j}\rangle\) are
Coulomb terms,
\(K_{ij} = \left\langle \phi_{i}\phi_{j} \middle| \phi_{j}\phi_{i} \right\rangle\)
are exchange terms, and \(V_{NN}\) is the nuclear repulsion energy. The
orbitals are determined self-consistently through the Fock equation,

\begin{equation}\label{eq:fock}
\widehat{F}\phi_i=\epsilon_i\phi_i,\qquad \widehat{F}=\widehat{h}+\sum_j\left(\widehat{J}_j-\widehat{K}_j\right).
\end{equation}

Although RHF recovers most of the total electronic energy, it neglects
correlation effects arising from instantaneous electron-electron
interactions beyond the mean-field picture. This missing contribution is
the correlation energy,

\begin{equation}\label{eq:correlation-energy}
E_{\mathrm{corr}} = E_{\mathrm{exact}} - E_{\mathrm{HF}}.
\end{equation}

which is negative.

For the open-shell radical fragments considered in this work, we employ
Restricted Open-Shell Hartree--Fock (ROHF) \cite{szabo1996modern}. ROHF preserves a
common spatial orbital description for doubly occupied orbitals while
allowing singly occupied orbitals, thereby avoiding the spin
contamination commonly associated with unrestricted Hartree--Fock. All
ROHF calculations are carried out with the Python Library PySCF
\cite{sun2018pyscf}.

\subsubsection{Coupled Cluster Singles and Doubles (CCSD)}\label{subsubsec:ccsd}

To recover dynamical electron correlation beyond the Hartree-Fock
reference, we use coupled cluster singles and doubles (CCSD) \cite{szabo1996modern}.
In CCSD, the correlated wavefunction is written as

\begin{equation}\label{eq:ccsd-wavefunction}
\ket{\Psi_{\mathrm{CCSD}}}=e^{\widehat{T}_1+\widehat{T}_2}\ket{\Phi_0}.
\end{equation}

where the cluster operators for single and double excitations are

\begin{equation}\label{eq:ccsd-cluster}
\begin{aligned}
\widehat{T}_1 &= \sum_{i \in \mathrm{occ}} \sum_{a \in \mathrm{virt}} t_i^a \, \hat{a}_a^{\dagger}\hat{a}_i, \\
\widehat{T}_2 &= \frac{1}{4}\sum_{ij}\sum_{ab} t_{ij}^{ab}\,\hat{a}_a^{\dagger}\hat{a}_b^{\dagger}\hat{a}_j\hat{a}_i.
\end{aligned}
\end{equation}

The amplitudes \(t_{i}^{a}\) and \(t_{ij}^{ab}\) are obtained by
projecting the similarity-transformed Schr\"{o}dinger equation onto the
singly and doubly excited determinants,

\begin{equation}\label{eq:ccsd-amplitudes}
\begin{aligned}
\bra{\Phi_i^a}\overline{H}\ket{\Phi_0} &= 0, \\
\bra{\Phi_{ij}^{ab}}\overline{H}\ket{\Phi_0} &= 0, \\
\overline{H} &= e^{-\widehat{T}}\widehat{H}e^{\widehat{T}}.
\end{aligned}
\end{equation}

The resulting CCSD correlation energy is

\begin{equation}\label{eq:ccsd-correlation}
E_{\mathrm{corr}}^{\mathrm{CCSD}}=\sum_{ia} t_i^a f_{ia}
+\frac{1}{4}\sum_{ijab}\left(t_{ij}^{ab}+t_i^a t_j^b-t_i^b t_j^a\right)\langle ij|ab\rangle.
\end{equation}

CCSD is size-extensive and scales as
\(\mathcal{O}\left( o^{2}v^{4} \right)\), where \(o\) and \(v\) denote
the number of occupied and virtual orbitals, respectively. In practice,
it captures the dominant fraction of the correlation energy and serves
here as the classical high-accuracy reference for fragment calculations.

\subsection{Quantum Methods}\label{subsec:quantum-methods}

\subsubsection{Variational Quantum Eigensolver (VQE)}\label{subsubsec:vqe}

The variational quantum eigensolver (VQE) \cite{peruzzo2014vqe} is used to estimate
fragment ground-state energies through a hybrid quantum-classical loop.
Given a parameterized ansatz \(U(\boldsymbol{\theta})\), the trial
state is

\begin{equation}\label{eq:vqe-state}
\ket{\psi(\boldsymbol{\theta})}=U(\boldsymbol{\theta})\ket{0}.
\end{equation}

and its energy is

\begin{equation}\label{eq:vqe-energy}
E(\boldsymbol{\theta})=\bra{\psi(\boldsymbol{\theta})}\widehat{H}\ket{\psi(\boldsymbol{\theta})}\ge E_0,
\end{equation}

where the inequality follows from the variational principle.

The workflow consists of three steps. First, the parameterized circuit
prepares the trial state on the quantum device. Second, the expectation
value of the Hamiltonian is estimated by measuring the Pauli terms in

\begin{equation}\label{eq:pauli-energy}
\widehat{H}=\sum_k c_k\widehat{P}_k,\qquad
E(\boldsymbol{\theta})=\sum_k c_k\expect{\widehat{P}_k}.
\end{equation}

Third, a classical optimizer updates \(\boldsymbol{\theta}\) to
minimize the measured energy.

In this work, we employ a hardware-efficient ansatz \cite{kandala2017hardware}, with
alternating layers of single-qubit \(R_{Y}\) and \(R_{Z}\) rotations
followed by \(CZ\) entangling gates in a linear topology. The circuit
depth is controlled by a hyperparameter and ranges from one to three
layers depending on the fragment.

Parameter optimization is carried out with SPSA (Simultaneous
Perturbation Stochastic Approximation) \cite{spall1998spsa}, which is particularly
suitable for noisy and shot-limited objective functions because it
estimates gradients using only two function evaluations per iteration,
independent of the number of parameters. Typical runs use 1500 to 2500
optimization steps and 10,000 shots per energy evaluation.

Because VQE is performed within the CASSCF active space constructed from
RHF or ROHF orbitals, the resulting energies provide variational upper
bounds to the corresponding active-space ground-state energies.

\subsubsection{Adaptive Derivative-Assembled Pseudo-Trotter VQE}\label{subsubsec:adapt-vqe}

To improve expressibility while avoiding unnecessarily large fixed
ans\"{a}tze, we also use ADAPT-VQE (Adaptive Derivative-Assembled
Pseudo-Trotter VQE) \cite{grimsley2019adapt}. In contrast to standard hardware-efficient
VQE, ADAPT-VQE iteratively grows the ansatz by selecting operators from
a physically motivated pool according to their instantaneous expected
contribution to the energy gradient. The algorithm starts from the
Hartree-Fock reference state \(\ket{\Phi_0}\). At each iteration,
the gradient associated with each operator \({\widehat{A}}_{k}\) in the
pool is evaluated as

\begin{equation}\label{eq:adapt-gradient}
g_k=\left.\left|\frac{\partial E}{\partial \theta_k}\right|\right|_{\theta_k=0}
=\left|\bra{\Phi}\left[\widehat{H},\widehat{A}_k\right]\ket{\Phi}\right|.
\end{equation}

The operator with the largest gradient
\(k^{*} = \arg{\max_{k}\left| g_{k} \right|}\) is selected, and the
corresponding unitary operator
\(e^{i\theta_{k^{*}}{\widehat{A}}_{k^{*}}}\) is appended to the circuit.
All parameters are then re-optimized simultaneously. This process
continues until the maximum gradient falls below a convergence
threshold, typically $10^{-3}$.

The operator pool used here consists of generalized singles and doubles
(GSD) excitations \cite{grimsley2019adapt,lee2019gucc}. The single-excitation generators are

\begin{equation}\label{eq:gsd-single}
\left(\widehat{G}\right)_{pq}^{(1)} = i\left(\hat{a}_p^{\dagger}\hat{a}_q - \hat{a}_q^{\dagger}\hat{a}_p\right).
\end{equation}

and the double-excitation generators are

\begin{equation}\label{eq:gsd-double}
\left(\widehat{G}\right)_{pqrs}^{(2)} = i\left(\hat{a}_p^{\dagger}\hat{a}_q^{\dagger}\hat{a}_r\hat{a}_s - \hat{a}_s^{\dagger}\hat{a}_r^{\dagger}\hat{a}_q\hat{a}_p\right).
\end{equation}

These anti-Hermitian operators guarantee unitary exponentials and allow
the ansatz to adapt to the structure of each fragment Hamiltonian.
Relative to fixed hardware-efficient circuits, ADAPT-VQE typically
achieves comparable or improved accuracy with fewer parameters and a
more compact, problem-informed circuit construction. It is also less
susceptible to barren plateau behavior than generic deep ans\"{a}tze
\cite{grimsley2023barren}.

\subsubsection{Sample-based Quantum Diagonalization (SQD)}\label{subsubsec:sqd}

Sample-based Quantum Diagonalization (SQD) \cite{robledoMoreno2025sqd} estimates
ground-state energies by combining quantum sampling with classical
subspace diagonalization. Instead of repeatedly measuring expectation
values during a variational optimization loop, SQD uses the quantum
device to generate bitstring samples from a trial state that overlaps
with the target ground state. These samples define a reduced
configuration subspace in which the Hamiltonian is projected and
diagonalized.

In this work, the sampled state is prepared with a circuit constructed
from CCSD amplitudes using a local unitary cluster Jastrow (LUCJ)
ansatz,

\begin{equation}\label{eq:lucj}
\ket{\Psi}=\prod_{\mu=1}^{L} e^{\widehat{K}_{\mu}}e^{i\widehat{J}_{\mu}}e^{-\widehat{K}_{\mu}}\ket{\Phi_0},
\end{equation}
where \(\ket{\Phi_0}\) is the Hartree--Fock reference. The circuit
is executed on IBM quantum hardware to produce computational-basis
bitstrings. To reduce noise, the measured samples are first
post-selected to preserve the correct
\(\left( n_{\alpha},n_{\beta} \right)\) sector. The surviving
configurations are then refined through the Self-Consistent Orbital
Recovery and Expansion (S-CORE) procedure \cite{robledoMoreno2025sqd}, which iteratively
improves the sampled subspace using estimated orbital occupancies.

Given the recovered configuration set \(\mathcal{S}\), the Hamiltonian
is projected as

\begin{equation}\label{eq:sqd-projection}
\widehat{H}_{\mathcal{S}} = \widehat{P}_{\mathcal{S}}\widehat{H}\widehat{P}_{\mathcal{S}},\qquad
\widehat{P}_{\mathcal{S}} = \sum_{x\in\mathcal{S}} \ket{x}\bra{x}.
\end{equation}
and the lowest eigenvalue of the projected Hamiltonian is taken as the
SQD energy estimate.

Unlike VQE, SQD avoids repeated quantum-classical optimization and
shifts most of the computational effort to classical post-processing.
This makes it attractive for near-term hardware, where noisy
measurements can often be filtered or corrected at the configuration
level. In this work, SQD complements VQE and ADAPT-VQE as a
sampling-based alternative for fragment energy estimation on IBM quantum
devices.

\section{Computational Details}\label{sec:computational-details}

\subsection{Geometry, Basis Set, and Benchmark Suite}\label{subsec:geometry}

All calculations are performed on linear alkanes $\mathrm{C}_n\mathrm{H}_{2n+2}$
from butane ($n=4$) to hexacosane ($n=26$), spanning 11 chain
lengths: $n=4,5,6,7,8,9,10,15,20,25,\text{ and }26$. Propane is excluded
because at $n=3$ the two terminal $\mathrm{CH}_3$ groups dominate the fragment sum,
producing boundary artifacts in the MBE2 decomposition that are not
representative of longer chains. Geometries are constructed in the
all-anti conformation with the carbon backbone in the \emph{xz}-plane,
using standard parameters
($(d_{\mathrm{CC}} = 1.54~\text{\AA},\ d_{\mathrm{CH}} = 1.09~\text{\AA},\ \angle\mathrm{CCC}=112^{\circ})$).
We use the STO-3G minimal basis set throughout, which assigns 5 atomic
orbitals per carbon and 1 per hydrogen. While STO-3G limits quantitative
accuracy, it provides a uniform framework for benchmarking and keeps
qubit counts within reach of current simulation and hardware
capabilities.

\subsection{Fragment Library and Classical References}\label{subsec:fragments}

Radical fragments are obtained by homolytic C--C bond cleavage of butane
as the prototype, yielding four unique fragment types that are reused
for all longer chains. The $\mathrm{CH}_3^{\bullet}$ monomer (methyl
radical) has 8 spatial orbitals and maps to 16 qubits under the
Jordan--Wigner transformation; it is a doublet state with
\(n_{\alpha}=5\) and \(n_{\beta}=4\) electrons (9 total). The
${}^{\bullet}\mathrm{CH}_2^{\bullet}$ monomer (methylene
diradical) has 7 spatial orbitals (14 qubits) with \(n_{\alpha}=5\)
and \(n_{\beta}=3\) electrons (8 total). The $\mathrm{CH}_3-\mathrm{CH}_2$
bonded dimer is the largest fragment at 15 spatial orbitals (30 qubits),
computed as a 17-electron doublet ($S=1/2$). The $\mathrm{CH}_2-\mathrm{CH}_2$
bonded dimer has 14 spatial orbitals (28 qubits) and is a 16-electron
triplet ($S=1$). All fragments are computed as isolated systems without
electrostatic embedding. For VQE and ADAPT-VQE, each fragment is further
reduced to a CAS($n$,4) active space yielding 8 qubits (2 $\times$ 4 active
spatial orbitals) under Jordan--Wigner mapping.

To anchor the fragmentation results, we compute full-molecule RHF and
CCSD energies for all 11 alkanes using PySCF \cite{sun2018pyscf}. These
unfragmented calculations serve two roles: RHF is the natural comparison
target for methods that operate on top of Hartree--Fock orbitals (VQE,
ADAPT-VQE), while CCSD is the reference for methods initialized from
correlated amplitudes (SQD). At the fragment level, ROHF and CCSD
energies are computed for each of the four fragment types, yielding the
MBE2-RHF and MBE2-CCSD baselines that isolate the fragmentation error
from the solver error. The CCSD $t_1$ and $t_2$ amplitudes and molecular
integrals are retained for initializing the quantum circuits described
below.

\subsection{Active-Space Reduction and Hamiltonian Construction}\label{subsec:active-space}

To reduce the calculation time, we compress each fragment into a uniform
8-qubit problem through a three-stage classical preprocessing pipeline.
First, ROHF provides mean-field orbitals for the open-shell radical.
Second, CASSCF selects a CAS($n$,4) active space (four spatial orbitals
and eight spin-orbitals) and simultaneously optimizes both the orbital
coefficients and the CI expansion within that space. The active-space
integrals (one-electron, two-electron, and the frozen-core energy)
extracted from CASSCF define the electronic problem that the quantum
solver must solve. Third, Cholesky decomposition \cite{aquilante2011cholesky} factorizes the
two-electron tensor to reduce the number of Pauli terms, and the
Jordan--Wigner transformation \cite{jordan1928pauli} maps the resulting fermionic
Hamiltonian to an 8-qubit Pauli operator. Before any variational
optimization begins, we verify by exact diagonalization that the qubit
Hamiltonian reproduces the CASSCF ground-state energy, ensuring that no
information is lost in the mapping. This active-space pipeline is applied to the VQE and ADAPT-VQE calculations, so those two solvers operate on uniform 8-qubit CAS(n,4) fragment Hamiltonians. SQD is treated separately: it is applied to the full STO-3G fragment Hamiltonians, requiring 14–30 qubits depending on the fragment.

\subsection{Quantum Solvers Configuration}\label{subsec:solver-config}

\subsubsection{VQE Configuration}\label{subsubsec:vqe-config}

Each fragment undergoes a pipeline to construct a qubit Hamiltonian in
the CAS($n$,4) active space (ROHF, CASSCF, Cholesky decomposition,
Jordan--Wigner mapping). The ansatz is an EfficientSU2 (NLocal) circuit
with alternating \(R_{Y}\)/\(R_{Z}\) single-qubit rotations and \(CZ\)
entangling gates in a linear connectivity pattern. The number of
repetition layers varies by fragment complexity: 1--2 layers for
monomers, 1--3 for dimers. Optimization uses SPSA \cite{spall1998spsa} with
1,500--2,500 maximum iterations and 10,000 shots per energy evaluation on
the Qiskit Aer statevector simulator. The CASSCF energy is verified by
exact diagonalization of the qubit Hamiltonian before VQE optimization.

\subsubsection{ADAPT-VQE Configuration}\label{subsubsec:adapt-config}

ADAPT-VQE uses the same CAS($n$,4) active space and Hartree-Fock initial
state as VQE. The operator pool consists of generalized singles and
doubles (GSD) \cite{grimsley2019adapt,lee2019gucc}, including \(\alpha \rightarrow \alpha\),
\(\beta \rightarrow \beta\), and mixed-spin double excitations across
all spin-orbital pairs. At each iteration, the energy gradient is
computed for every pool operator via the commutator
\(\langle\Psi|\lbrack H,\ A_{k}\rbrack|\Psi\rangle\), and the operator
with the largest gradient magnitude is appended to the ansatz. All
parameters are re-optimized after each addition. Convergence is declared
when the maximum gradient falls below $10^{-3}$. The statevector
simulator is used for gradient evaluation and energy optimization.

\subsubsection{SQD Configuration and IBM Quantum Hardware}\label{subsubsec:sqd-config}

SQD circuits are constructed from the CCSD $t_1$ and $t_2$ amplitudes
extracted from each fragment using the LUCJ ansatz \cite{robledoMoreno2025sqd}. The orbital
rotation parameters are initialized from $t_1$ and the Jastrow parameters
from $t_2$. Each circuit is transpiled for the target IBM quantum backend
and executed with 25,000 measurement shots via the Qiskit SamplerV2
primitive. Raw bitstring counts are saved to JSON files for
reproducibility. The classical post-processing uses 10 self-consistent
S-CORE iterations. At each iteration, the recovered configurations are
partitioned into 10 independent batches of 2,500 samples, and the
Hamiltonian is diagonalized in each batch subspace using the Davidson
iterative method with up to 400 cycles. The lowest eigenvalue across all
batches and iterations is taken as the fragment SQD energy.

\section{Results and Analysis}\label{sec:results}

To differentiate the error introduced by fragmentation from the error
introduced by the quantum solver, we adopt a two-layer comparison
strategy. First, we compare each MBE2 reconstruction against the
corresponding unfragmented calculation (MBE2-RHF vs full RHF, MBE2-CCSD
vs full CCSD) to quantify the intrinsic cost of the bonded-pair-only
decomposition. Second, we separate the quantum-solver contribution from the active-space and fragmentation contributions. For VQE and ADAPT-VQE, the appropriate solver reference is exact diagonalization of the same CAS(n,4) qubit Hamiltonians used by the quantum algorithms. Comparisons to MBE2-RHF are retained only as chemically useful baseline comparisons, not as pure solver-error estimates. For SQD, which is applied to the full fragment Hamiltonians, MBE2-CCSD is used as a practical correlated classical reference, while the SQD result should be interpreted as a hardware-sampled subspace-diagonalization estimate rather than a direct approximation to CCSD. This separation is essential
because, as the results below show, the fragmentation approximation
dominates the total error budget for every method. All energies are
reported in Hartree (Ha) and errors in kcal/mol (1 Ha = 627.509
kcal/mol).

\subsection{Intrinsic Fragmentation Error}\label{subsec:fragmentation-error}

\begin{figure}[!htbp]
    \centering
    \includegraphics[width=0.9\linewidth]{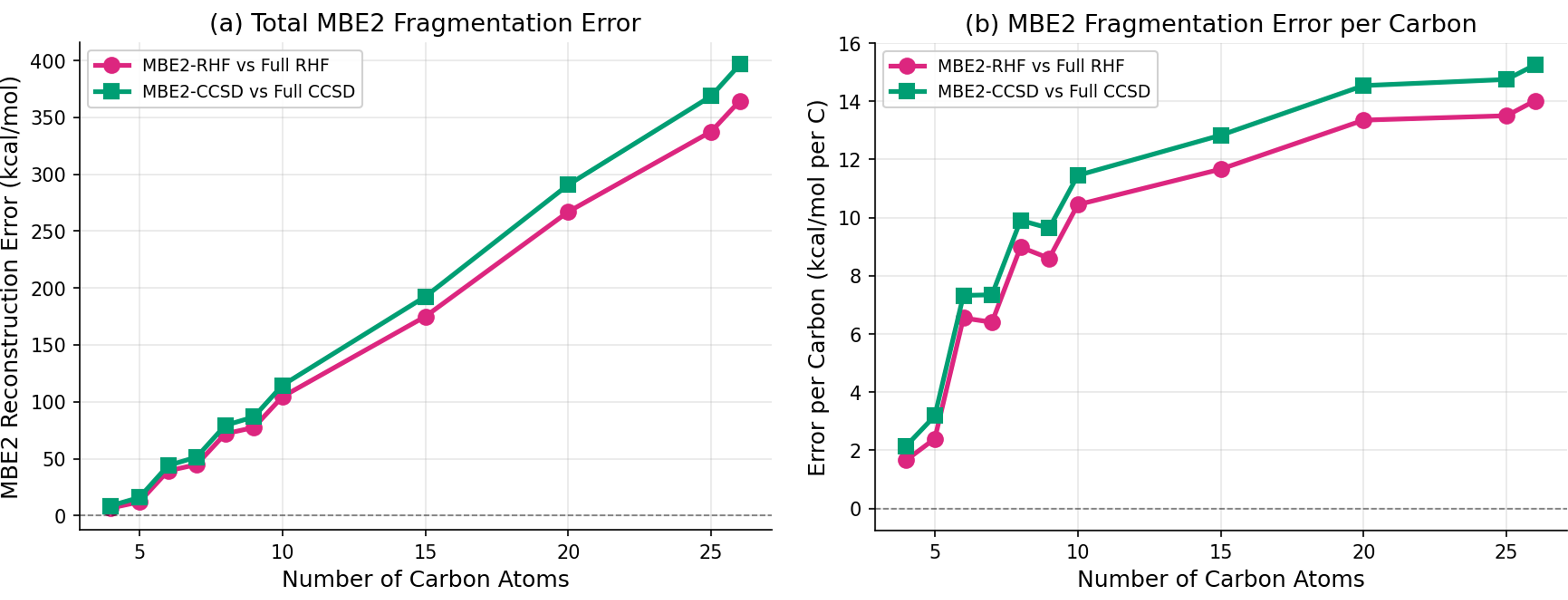}
    \caption{Intrinsic MBE2 fragmentation error for the two classical methods. (a) Total reconstruction error (kcal/mol) vs. number of carbon atoms for MBE2-RHF (relative to full RHF) and MBE2-CCSD (relative to full CCSD). (b) Error normalized per carbon atom, showing convergence toward a stable per-carbon penalty as chain length increases.}
    \label{fig:fragmentation-error}
\end{figure}

The error baseline presented in Fig.~\ref{fig:fragmentation-error} is established by the energy
penalty of decomposing a full molecule into radical fragments with only
bonded-pair corrections. MBE2-RHF overshoots full-molecule RHF by +6.6
kcal/mol for butane and +364.6 kcal/mol for hexacosane; MBE2-CCSD shows
a parallel trend at +8.6 and +396.5 kcal/mol, respectively. The slightly
larger CCSD fragmentation error is expected: correlation energy has
longer-range character than the mean-field energy, so truncating the
expansion at bonded pairs discards a larger fraction of the CCSD
correction. Both curves grow with chain length but the error per carbon
atom converges from +1.7 to +14.0 kcal/mol per C for MBE2-RHF and from
+2.2 to +15.3 kcal/mol per C for MBE2-CCSD, confirming that the MBE2
per-carbon error converges to a well-defined asymptotic value,
consistent with the linear (size-extensive) scaling expected from the
constant per-$\mathrm{CH}_2$ increment inherent in the assembly formula. The
close tracking of the two curves across all 11 alkanes demonstrates that
the fragmentation error is structural, arising from neglected non-bonded
pairs and three-body terms.

\subsection{Quantum Solver Accuracy}\label{subsec:solver-accuracy}

\begin{figure}[!htbp]
    \centering
    \includegraphics[width=0.9\linewidth]{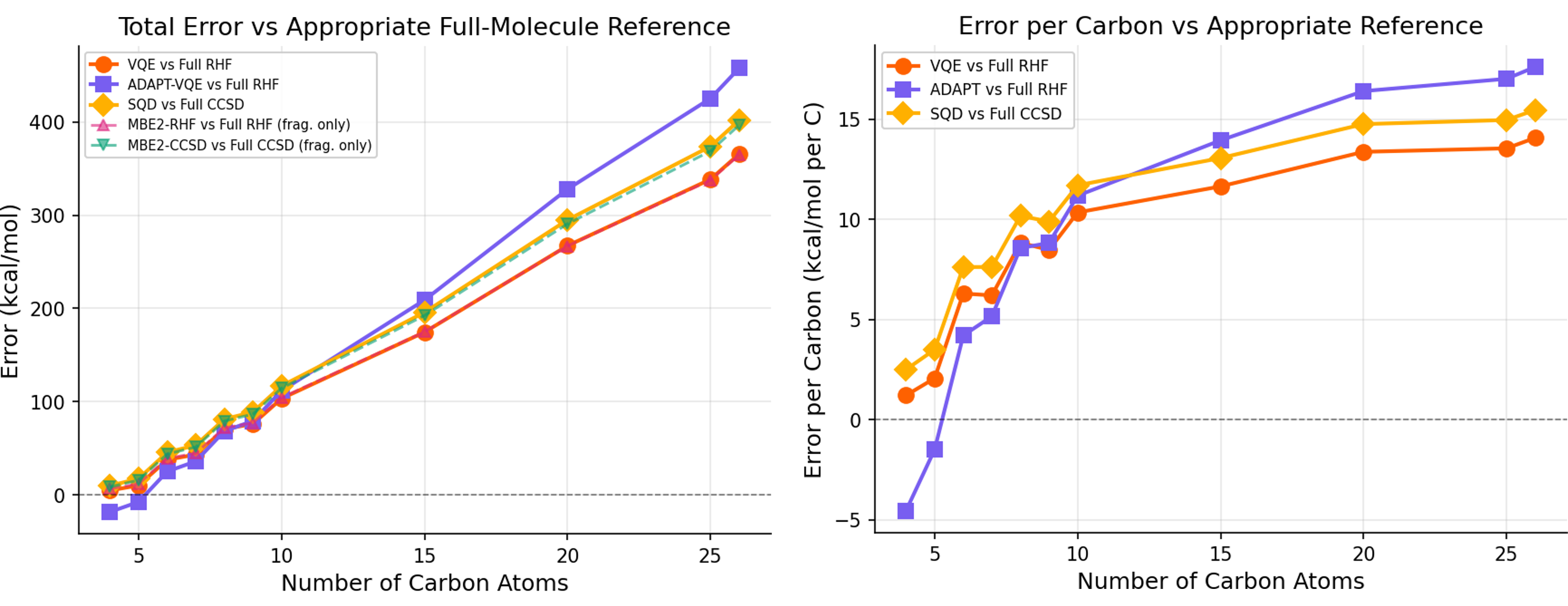}
    \caption{Comprehensive error comparison. (a) Total error (kcal/mol) for all three quantum methods against their respective full-molecule references, overlaid with the pure fragmentation baselines MBE2-RHF and MBE2-CCSD. (b) Error per carbon atom, showing convergence toward stable per-carbon asymptotes for long chains.}
    \label{fig:solver-error}
\end{figure}

The most informative comparison is between each quantum solver and its
own MBE2-classical reference. This solver-only error reveals how much
accuracy is lost by replacing the classical eigensolver with a quantum
algorithm. For VQE, the total error relative to full-molecule RHF ranges
from +4.9 kcal/mol for butane to +365.6 kcal/mol for hexacosane, closely
paralleling the MBE2-RHF fragmentation error across the entire series.
The near-coincidence of the two curves in Fig.~\ref{fig:solver-error} (orange) confirms that
VQE adequately recovers the active-space energy within each fragment,
the gap between MBE2-VQE and MBE2-RHF remains small and stable, never
exceeding a few kcal/mol, while the total error is dominated by the
bonded-pair-only fragmentation approximation. In relative terms, the
MBE2-VQE energy deviates from full RHF by only 0.005\% for butane and
0.058\% for hexacosane. The per-carbon error grows from +1.2 to +14.1
kcal/mol per C, converging toward the same asymptote as MBE2-RHF.

ADAPT-VQE also exhibits a chain-length-dependent error trend. Both VQE
and ADAPT-VQE solve the same CASSCF active-space Hamiltonians, but they
differ in optimization strategy: VQE uses a fixed EfficientSU2 ansatz
with shot-based SPSA, while ADAPT-VQE grows its circuit from the GSD
pool using exact statevector evaluation and a tight gradient threshold
of $10^{-3}$. ADAPT-VQE therefore converges closer to the exact CASSCF
ground state for each fragment, producing lower fragment energies. At
butane, the assembled ADAPT energy lies 24.8 kcal/mol below MBE2-RHF,
compared to only 1.8 kcal/mol for VQE. However, this tighter
per-fragment convergence is unbalanced between monomers and dimers: the
per-$\mathrm{CH}_2$ increment (encoding the difference
$E_{\mathrm{CH}_2-\mathrm{CH}_2}-E_{\mathrm{CH}_2}$) deviates +5.35 kcal/mol from the ROHF
baseline for ADAPT-VQE, versus only +0.13 kcal/mol for VQE. The MBE2
assembly formula multiplies this monomer--dimer imbalance by
$(n-3)$, so it accumulates linearly with chain length. For
butane (one $\mathrm{CH}_2-\mathrm{CH}_2$ pair correction), the tighter convergence
still dominates and the total error is -18.1 kcal/mol relative to full
RHF. By hexacosane (with 23 $\mathrm{CH}_2-\mathrm{CH}_2$ corrections), the accumulated
imbalance reaches +457.5 kcal/mol. VQE avoids this drift not because it
is a better solver, but because its shot-based optimization converges
less tightly and more uniformly across fragment types, producing a
per-$\mathrm{CH}_2$ increment that nearly matches the ROHF reference.

Unlike VQE and ADAPT-VQE, which solve reduced 8-qubit CASSCF
Hamiltonians on a statevector simulator, SQD operates on the full
fragment Hamiltonians: 16 qubits for $\mathrm{CH}_3^{\bullet}$, 14 for
${}^{\bullet}\mathrm{CH}_2^{\bullet}$, 30 for the $\mathrm{CH}_3-\mathrm{CH}_2$ dimer, and
28 for $\mathrm{CH}_2-\mathrm{CH}_2$ executed on IBM's \texttt{ibm\_pittsburgh} ,
which is one of the IBM Quantum Heron Processors, with 25,000 shots per
fragment. Because SQD works in the same full orbital space as CCSD and
is initialized from CCSD amplitudes, the natural comparison target is
MBE2-CCSD. The total error relative to full-molecule CCSD ranges from
+10.0 kcal/mol (butane) to +401.7 kcal/mol (hexacosane), closely
tracking the MBE2-CCSD fragmentation error of +8.6 to +396.5 kcal/mol.
The gap between the two is only +1.4 kcal/mol for butane and +5.2
kcal/mol for hexacosane, growing at +0.17 kcal/mol per additional $\mathrm{CH}_2$
unit. This per-$\mathrm{CH}_2$ deviation is comparable to VQE's
+0.13 kcal/mol despite SQD solving Hamiltonians 2--4$\times$ larger on real
hardware rather than an ideal simulator.

When each quantum method is compared to the best available unfragmented
calculation (VQE and ADAPT-VQE vs full RHF; SQD vs full CCSD), the total
error naturally includes both fragmentation and solver contributions.
Fig.~\ref{fig:solver-error} displays all five error curves, including the pure fragmentation
baselines MBE2-RHF and MBE2-CCSD. The per-carbon error (Fig.~\ref{fig:solver-error}, bottom)
shows convergence for all methods, with the long-chain asymptotes being
+14.1 kcal/mol per C (VQE vs RHF), +15.4 kcal/mol per C (SQD vs CCSD),
and +17.6 kcal/mol per C (ADAPT-VQE vs RHF).

\subsection{Qubit Reduction and Computational Scaling}\label{subsec:qubit-scaling}

\begin{figure}[H]
    \centering
    \includegraphics[width=0.58\linewidth]{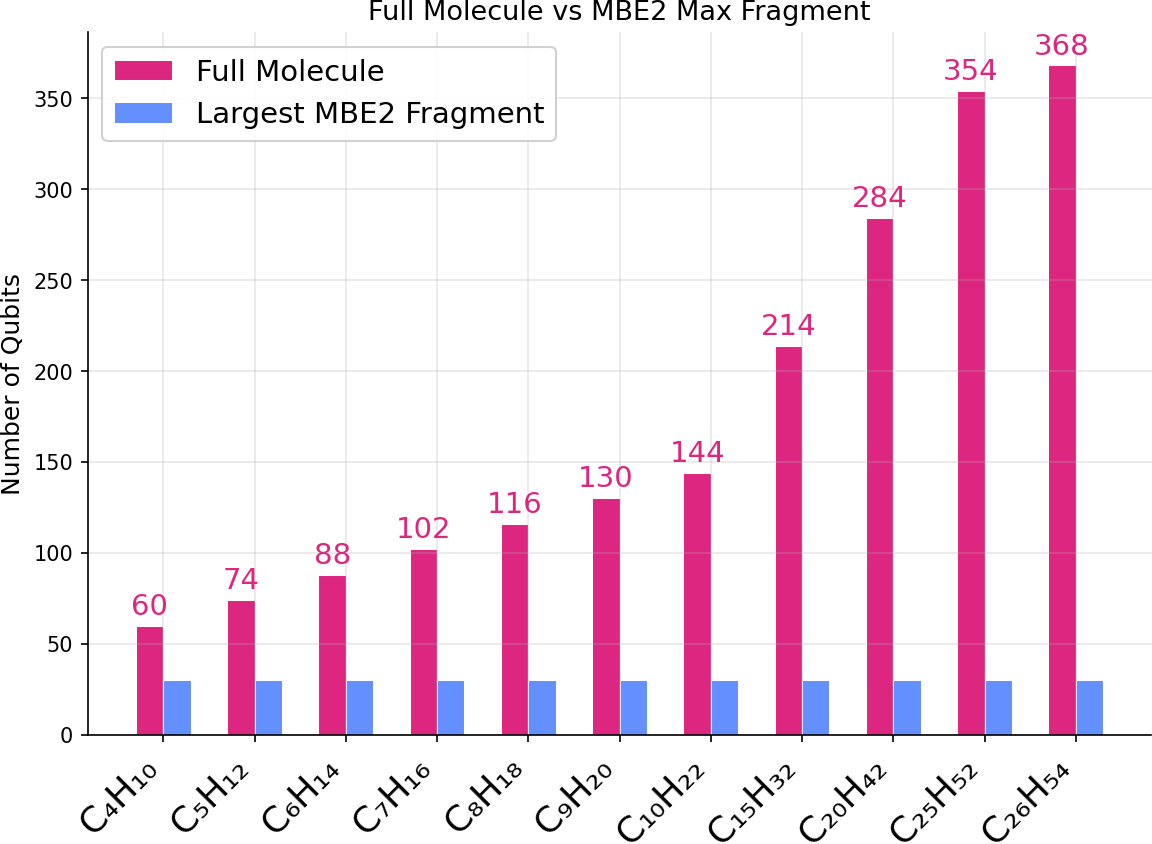}
    \caption{Qubit requirements. Side-by-side bar chart of full-molecule qubit count vs. the constant MBE2 maximum fragment (30 qubits) for each alkane.}
    \label{fig:qubit-requirements}
\end{figure}

The practical motivation for the entire fragmentation approach is the
qubit savings summarized in Fig.~\ref{fig:qubit-requirements}. The
full-molecule bars grow linearly from 60 (butane) to 368 (hexacosane),
while the MBE2 bars remain fixed at 30 qubits.

\begin{figure}[H]
    \centering
    \includegraphics[width=0.90\linewidth]{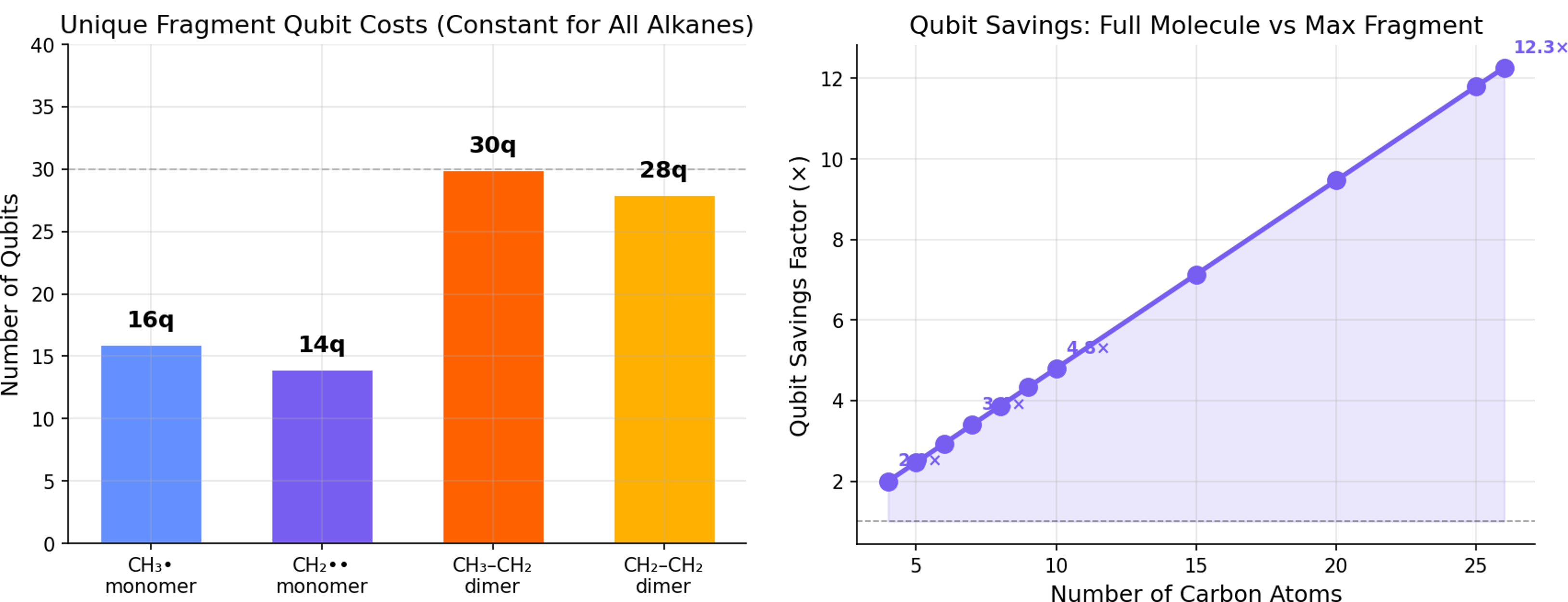}
    \caption{
    Fragment qubit costs and qubit-savings factor for the radical-fragment MBE2 construction.
    (a) Qubit count for each unique full-fragment Hamiltonian in the STO-3G basis:
    CH$_3^\bullet$ monomer, 16 qubits;
    $^\bullet$CH$_2^\bullet$ monomer, 14 qubits;
    CH$_3$--CH$_2$ dimer, 30 qubits;
    and CH$_2$--CH$_2$ dimer, 28 qubits.
    (b) Qubit-savings factor obtained by comparing the full-molecule Jordan--Wigner/STO-3G qubit count,
    $q = 14n + 4$, with the maximum full-fragment requirement of 30 qubits.
    For C$_{26}$H$_{54}$, this gives a reduction from 368 qubits to 30 qubits,
    corresponding to a savings factor of $368/30 \approx 12.3$.
    }
    \label{fig:fragment-qubit-costs}
\end{figure}

In Fig.~\ref{fig:fragment-qubit-costs}, the lower panel recasts these qubit savings into a scaling
diagram, with the shaded region between the linear full-molecule curve
($q\approx 14n+4$) and the flat MBE2 line representing the growing qubit
savings. The savings factor rises from 2.0$\times$ for butane through 4.8$\times$ for
decane and 9.5$\times$ for eicosane to 12.3$\times$ for hexacosane, a 92\% qubit
reduction at the largest chain. For the VQE and ADAPT-VQE active-space
calculations, the effective qubit count is further reduced to 8 qubits
per fragment, making the quantum computations feasible on any current
device.

The symmetry of linear alkanes compounds this advantage. Although
hexacosane decomposes into 26 monomers and 25 dimers (51 fragments
total), only 4 are treated as chemically unique. This 12.3$\times$ reduction in
the number of quantum calculations is independent of the qubit savings.
Fig.~\ref{fig:fragment-qubit-costs} presents the growing gap between brute-force fragment counts and
the constant 4 unique calculations, alongside the qubit savings factor.
Together, the qubit ceiling and the constant calculation count mean that
extending the radical-fragment MBE2 approach to arbitrarily long alkane
chains requires no additional quantum resources whatsoever.

\FloatBarrier
\section{Discussion}\label{sec:discussion}

The results demonstrate that radical-fragment MBE2 is a viable strategy
for extending quantum chemistry solvers to large molecular systems. The
constant fragment size ensures that the quantum hardware requirements do
not grow with the target molecule, while exploiting symmetry reduces the
number of quantum computations required.

Among the three quantum solvers, SQD stands out for its practical
advantages on near-term hardware. By treating the quantum processor as a
sampling oracle rather than an energy evaluation engine, SQD avoids the
measurement overhead and noise sensitivity that constrain VQE. The LUCJ
circuit initialized from CCSD amplitudes provides sufficient overlap
with the ground state for the configuration recovery algorithm to
extract accurate energies even from noisy hardware samples.

The dominant source of error in all MBE2 methods is the bonded-pair-only
approximation rather than the quantum solver itself. This is evidenced
by the similar growth patterns of MBE2-RHF, MBE2-CCSD, and MBE2-SQD
errors with chain length. Including non-bonded pair corrections or
three-body (MBE3) terms would improve accuracy at the cost of additional
fragment calculations, though the number of unique calculations would
remain bounded by symmetry.

The error per carbon atom converges for long chains in all methods,
confirming approximate size-extensivity. This property is essential for
thermodynamic predictions and suggests that the framework would maintain
accuracy for arbitrarily long alkane chains.

\section{Conclusions}\label{sec:conclusions}

We have introduced a radical-fragment many-body expansion (MBE2)
approach that makes quantum chemistry of large molecules tractable on
near-term quantum hardware. By performing homolytic C--C bond cleavage to
produce open-shell radical fragments computed via ROHF without
electrostatic embedding or capping atoms, the method reduces the quantum
resource requirements from $O(n)$ to $O(1)$: the largest fragment requires
only 30 qubits regardless of chain length, compared to the 368 qubits
needed for a full-molecule treatment of hexacosane under JW/STO-3G,
representing a 12.3$\times$ qubit reduction.

The benchmark across three quantum solvers confirms that this
fragmentation framework faithfully preserves the accuracy of the
underlying solver. MBE2-VQE tracks the MBE2-RHF baseline with a per-$\mathrm{CH}_2$
deviation of only +0.13 kcal/mol, demonstrating that the 8-qubit CASSCF
Hamiltonians produced by the fragmentation pipeline are well-suited to
hardware-efficient ansatz optimization. MBE2-SQD goes further: operating
on the full fragment Hamiltonians (14-30 qubits) on
IBM's \texttt{ibm\_pittsburgh} processor, it reproduces MBE2-CCSD
energies with a hardware penalty of only +1.4 to +5.2 kcal/mol across
the entire series, evidence that the radical-fragment decomposition
pairs naturally with sample-based quantum algorithms on current
hardware. MBE2-ADAPT-VQE reveals that the MBE2 assembly step is
sensitive to how evenly the monomer and dimer fragment energies are
converged: if one fragment type is solved to higher accuracy than the
other, the energy subtraction inherent in the many-body expansion
amplifies that imbalance, producing larger errors in the total energy.
This observation highlights that, in future applications, the choice of
quantum solver and the convergence thresholds applied to each fragment
must be carefully matched to maintain a balanced level of accuracy
across all fragment types.

The accuracy of the assembled energies is currently bounded by the
bonded-pair-only approximation, which is a known and systematically
improvable limitation: incorporating non-bonded pair corrections and
MBE3 three-body terms within the same radical-fragment framework would
tighten the fragmentation baseline without increasing the maximum
fragment size. Future work will pursue these extensions alongside
application to branched hydrocarbons, heteroatom-containing systems, and
larger basis sets, regimes where the constant-size fragment advantage of
the radical MBE2 approach becomes even more decisive. As current
hardware qubit limitations are lifted in the future, this work can be
gradually extended to other types of polymer systems with increasingly
larger levels of electron delocalization, eventually allowing
calculation of highly delocalized electron systems, such as conductive
polymers.

\section*{Acknowledgments}\label{sec:acknowledgments}

We acknowledge the use of IBM Quantum Credits for this work. The views
expressed are those of the authors, and do not reflect the official
policy or position of IBM or the IBM Quantum team.

\section*{Disclaimer}\label{sec:disclaimer}

This work has been submitted to the IEEE for possible publication.
Copyright may be transferred without notice, after which this version
may no longer be accessible.

\end{document}